\documentclass[a4paper,conference]{IEEEtran}
\IEEEoverridecommandlockouts
\usepackage{cite}
\usepackage[hyphens]{url}
\usepackage{breakurl}
\usepackage{mathtools}
\usepackage{amsmath,amssymb,amsfonts}
\usepackage{algorithmic}
\usepackage{graphicx}
\usepackage{textcomp}
\usepackage{xcolor}
\usepackage{url}
\usepackage{footmisc}

\def\BibTeX{{\rm B\kern-.05em{\sc i\kern-.025em b}\kern-.08em
    T\kern-.1667em\lower.7ex\hbox{E}\kern-.125emX}}
\begin{document}

\title{Non-Linear Self Augmentation Deep Pipeline for Cancer Treatment outcome Prediction \\

}

\author{\IEEEauthorblockN{Francesco Rundo}
\IEEEauthorblockA{\textit{ADG Central R\&D} \\
\textit{STMicroelectronics}\\
Catania, Italy \\
francesco.rundo@st.com}
\and
\IEEEauthorblockN{Concetto Spampinato}
\IEEEauthorblockA{\textit{PerCeiVe Lab} \\
\textit{University of Catania}\\
Catania, Italy \\
cspampinato@dieii.unict.it}
\and
\IEEEauthorblockN{Michael Rundo}
\IEEEauthorblockA{\textit{PerCeiVe Lab} \\
\textit{University of Catania}\\
Catania, Italy \\
michael.rundo3764@gmail.com}
}

\maketitle

\begin{abstract}
Immunotherapy emerges as 
promising approach for treating cancer. Encouraging findings have
validated the efficacy of immunotherapy medications in addressing tumors, resulting in prolonged survival rates and notable
reductions in toxicity compared to conventional chemotherapy
methods. However, the pool of eligible patients for immunotherapy remains relatively small, indicating a lack of comprehensive
understanding regarding the physiological mechanisms responsible for favorable treatment response in certain individuals while
others experience limited benefits. To tackle this issue, the authors
present an innovative strategy that harnesses a non-linear cellular
architecture in conjunction with a deep downstream classifier.
This approach aims to carefully select and enhance 2D features
extracted from chest-abdomen CT images, thereby improving
the prediction of treatment outcomes. The proposed pipeline
has been meticulously designed to seamlessly integrate with an
advanced embedded Point of Care system. In this context, the
authors present a compelling case study focused on Metastatic
Urothelial Carcinoma (mUC), a particularly aggressive form
of cancer. Performance evaluation of the proposed approach
underscores its effectiveness, with an impressive overall accuracy
of approximately 93\%
\end{abstract}

\begin{IEEEkeywords}
Cellular Non-Linear Network, Deep Convolutional Network, Immunotherapy, Radiomics
\end{IEEEkeywords}

\section{Introduction}
Immunotherapy has emerged as the forefront in the battle against cancer, revolutionizing the approach by treating tumors as if they were infectious agents. The concept involves empowering the patient's immune system to recognize and eliminate cancer cells, akin to arming the body's defenses \cite{alsaab2017pd, spencer2016biomarkers}. While the immune system naturally recognizes cancer cells and activates T lymphocytes to attack them, cancer cells possess mechanisms to evade this response, exploiting the immune system's self-regulation process through various proteins acting as "accelerators" or "brakes" on T cells \cite{alsaab2017pd, spencer2016biomarkers, ding2019clinicopathological, zhou2017review}.
A promising strategy in immunotherapy focuses on inhibiting "Immunological Checkpoints" (ICIs) by employing specific antibodies. These ICIs counteract the disabling effects imposed by cancer cells, thereby boosting the efficacy of T lymphocytes in combating tumors \cite{alsaab2017pd, spencer2016biomarkers, ding2019clinicopathological}. In this context, our discussion will revolve around immunotherapy treatments targeting the PD-1 receptor, which represent a significant advancement in the field \cite{ding2019clinicopathological}.
Research studies have revealed a crucial mechanism employed by cancer cells to evade the immune system's response. They utilize a molecule called PD-L1, which is present on their membrane and binds to the PD-1 receptor on T lymphocytes, effectively disabling their protective actions. In the context of immunotherapy, ICIs targeting the PD-1/PD-L1 interaction aim to inhibit these receptors, enabling T lymphocytes to recognize and eliminate cancer cells \cite{ding2019clinicopathological, zhou2017review}.
In this study, we investigate this mechanism specifically in the case of metastatic bladder cancer, focusing on metastatic Urothelial Carcinoma (mUC) \cite{zhou2017review}. Urothelial carcinoma is known to be one of the most aggressive forms of cancer affecting the urinary system, leading to a significant number of deaths annually \cite{alsaab2017pd, ferlay2015cancer}. Platinum-based chemotherapy is currently considered the standard treatment for mUC, as demonstrated by various clinical trials \cite{zhou2017review}. However, understanding the dynamics of long-term survival rates in patients with bladder cancer undergoing chemotherapy or immunotherapy is crucial, and it is assessed through parameters such as Progression-Free Survival (PFS) rate and Overall Survival (OS) rate, which provide quantitative measures of treatment effectiveness \cite{de2012randomized}.
In the case of high-dose chemotherapy treatment, the median Overall Survival (OS) ranges from 12 to 15 months for a cisplatin-based regimen and approximately 9 months for a carboplatin-based regimen \cite{seront2015molecular}. However, these chemotherapy treatments are associated with significant toxicity. In contrast, immunotherapy has emerged as a new standard of care due to its potential benefits. Immune checkpoint inhibitors (ICIs) such as Atezolizumab and Pembrolizumab have shown a median OS of more than 10 months in both cases, with a notable reduction in side effects compared to chemotherapy treatments \cite{powles2018atezolizumab, bellmunt2017pembrolizumab}.
Despite the effectiveness of immunotherapy, only a subset of patients, approximately 20\% to 30\%, experience a positive response \cite{sharma2016efficacy, massard2016safety, apolo2017avelumab}. Therefore, the development of discriminative biomarkers capable of identifying patients who are likely to benefit from immunotherapy is a primary focus of current research activities.
In recent years, there has been a growing interest in using visual features from medical images and biomarkers such as PD-L1 expression level or blood indexes like the neutrophil to lymphocyte ratio as predictive immunotherapy outcome biomarkers \cite{banna2019promise}. The field of Radiomics, which involves the quantitative analysis of large sets of biomedical multi-modal data, has played a significant role in linking clinical trials to imaging research \cite{lambin2012radiomics, wang2015prediction, garapati2017urinary}.
In this study, we propose an innovative and less invasive pipeline for bladder cancer diagnosis based on deep learning algorithms. Specifically, we focus on classifying visual features extracted from chest-abdomen CT-scan images using a specially configured Cellular Non-Linear Network (CNN). This pipeline aims to accurately identify and augment CT cancer lesions. Furthermore, our designed pipeline is intended to be implemented on an embedded hardware accelerated Point of Care system, ensuring its practicality and efficiency.

\section{Related Works}

Accurate prediction of medical treatment outcomes for metastatic diseases requires the development of robust quantitative data processing pipelines. Machine and Deep Learning approaches have been extensively explored in the scientific literature for this purpose \cite{wang2015prediction, garapati2017urinary, hasnain2019machine}.
In a study by Wang et al. \cite{wang2015prediction}, the performance of various Machine Learning (ML) algorithms was evaluated to predict mortality after radical cystectomy in bladder cancer patients. The Regularized Extreme Learning Machine demonstrated superior accuracy compared to other methods. Similarly, Garapati et al. \cite{garapati2017urinary} applied common ML algorithms to analyze CT-scan urography in a clinical study, with Support Vector Machine (SVM) achieving impressive results.
Hasnain et al. \cite{hasnain2019machine} proposed ML methods, including K-Nearest Neighbors, AdaBoost, and SVM, for estimating bladder cancer disease. Their approach focused on predicting cancer recurrence and survival using multi-modal data analysis. The results showed specificity and sensitivity higher than 70\%, highlighting the effectiveness of Radiomics in medical oncology.
The advent of large-scale multi-source medical data has transformed the learning model paradigm, leading to more efficient data processing approaches. Various solutions for cancer image-lesion segmentation, including 2D and 3D convolutional networks, have been proposed \cite{hasnain2019machine, cha2016urinary, gordon2017segmentation, ma20192d, shkolyar2019augmented}. These methods have shown promising performance in segmenting metastatic lesions from CT imaging, particularly when dealing with lesions of the same type (visceral or lymphatic).

Additional deep pipelines for estimating the response to cancer treatments based on quantitative data analysis have been proposed in recent research \cite{wu2019deep, krizhevsky2012imagenet, cha2017bladder}.
In \cite{wu2019deep}, a modified version of the AlexNet backbone \cite{krizhevsky2012imagenet} was employed to develop a Deep Learning architecture for assessing the response to chemotherapeutic treatment. The proposed deep network was trained on segmented CT slices to learn visual features and achieved promising results in predicting treatment outcomes.
Cha et al. \cite{cha2017bladder} presented a Deep Learning pipeline for predicting immunotherapy outcomes in bladder cancer patients. The pipeline consisted of two stages: primary cancer-lesion segmentation in CT scans and treatment response prediction using a deep architecture. The performance of the pipeline was evaluated on a dataset of 82 bladder cancer patients, achieving a specificity of 81\% (DL-CNN) and a sensitivity of 66.7\% (RF-ROI).
In \cite{rundo2019advanced}, the authors introduced a novel deep pipeline for detecting immunotherapy outcomes in bladder cancer patients using artificial intelligence technology. The pipeline involved a deep classifier trained on visual features generated by a stack of encoders. The results showed improved performance compared to the authors' previous solution, with an accuracy of 86.05\%, specificity of 89.29\%, and sensitivity of 80.00\%.
These advanced deep pipelines demonstrate the progress made in using deep learning and artificial intelligence techniques to predict treatment outcomes in bladder cancer patients, offering promising results and potential for embedded systems in clinical practice \cite{wu2019deep, cha2017bladder, rundo2019advanced}.

 \begin{figure*}

  \centering
  \centerline{\includegraphics[width=\linewidth]{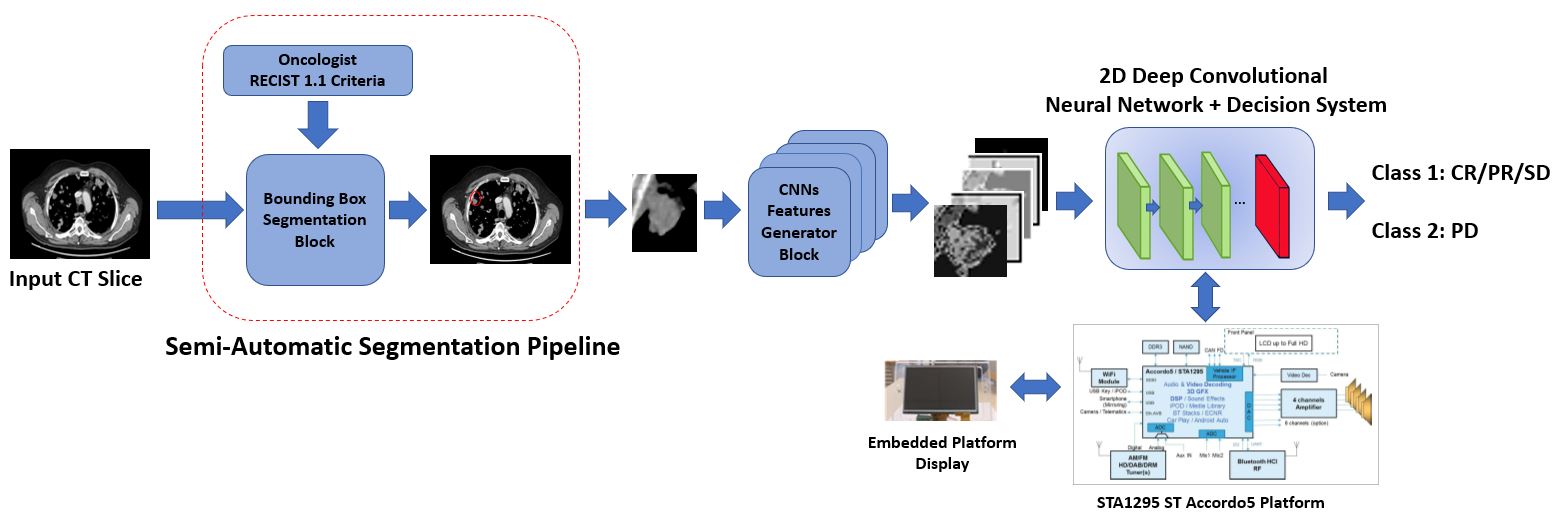}}
  \caption{\small The proposed immunotherapy treatment outcome prediction pipeline. }
  \label{fig:Img1} 
\end{figure*}

\section{The proposed Deep Network Framework}

 In Fig. \ref{fig:Img1}, we reported the comprehensive architecture of the proposed deep pipeline, designed to predict the immunotherapy treatment outcome for patients diagnosed with metastatic bladder cancer. The pipeline leverages radiomics techniques to quantitatively analyze selected chest-abdomen CT-scan imaging cancer lesions. The identification of CT-scan lesions is carried out by experienced oncologists/radiologists in accordance with the RECIST 1.1 guideline \cite{eisenhauer2009new}.
Unlike many existing pipelines \cite{cha2017bladder}, the proposed approach takes into account both primary lesion images (bladder) and metastatic lesion images (visceral or lymph nodes). This unique feature enhances the method's robustness and efficiency in the medical domain.
Initially, the semi-automatic pipeline performs a bounding box segmentation of the CT lesion identified by skilled physicians. This segmentation defines the Region of Interest (ROI) to which the predictive pipeline is applied. Those ROI images have different biologically visual features as they have been extracted from various body-sites of mUC metastasis (lungs, abdominal organs, bladder, lymph nodes, etc ..). The diverse nature of the dataset poses a challenge in constructing a fully automated semantic segmentation pipeline. Consequently, the segmented ROIs, representing the CT lesions, undergo further processing using a customized and expanded 2D Cellular Non-Linear Network (2D-CNN). This network generates a set of augmented domain-agnostic features, which are then fed into a downstream 2D Deep Classifier (2D-DNN). The purpose of the 2D-DNN is to accurately classify the augmented visual features and predict the treatment response of patients. Specifically, it distinguishes patients who may exhibit a favorable response to immunotherapy treatment (CR: Complete Response / PR: Partial Response / SD: Stable Disease) from those who experience disease progression (PD: Progressive Disease). Several 2D-DNN backbones have been tested preferring deep architectures with medium-low complexity to make the overall pipeline portable on embedded platforms based on STA1295 ASIL-B certified hardware architecture\footnote{\url{https://www.st.com/en/automotive-infotainment-and-telematics/sta1295.html}\label{note1}}. The following sections provide a detailed description of the proposed solution.

\subsection{The Bounding Box Segmentation block}

The proposed pipeline incorporates a semi-automatic CT-scan lesion segmentation process guided by experienced oncologists/radiologists. Initially, the entire chest-abdomen CT scan of each patient is examined, and the oncologist/radiologist manually identifies a specific cancerous lesion based on the criteria outlined in the RECIST revision 1.1 guideline \cite{eisenhauer2009new}. Most modern CT imaging software includes automated ROI selection based on spatial, dimensional, or morphological criteria. Following the lesion selection, our segmentation software automatically extracts a bounding box ROI of size $M \times N$ centered around the lesion. The specific dimensions (M, N) can vary, and the overall performance of the pipeline remains unaffected as the segmented lesion will be subsequently rescaled to match the input size of the downstream 2D-DNN classifier. For this particular implementation, we opted for a bounding box ROI size of $64 \times 64$. Additional details regarding the RECIST 1.1 guideline can be found in \cite{eisenhauer2009new}. The term RECIST refers to an official standardized methodology for evaluating treatment outcomes in solid tumors \cite{eisenhauer2009new}. The RECIST criteria, as of interest in this research work, includes the following items:
\begin{itemize}
\item Measurable lesion (target lesion): a lesion that can be accurately measured with longest diameter (in one dimension) $\geq$ 20 mm (CT scan imaging);
\item A sum of the longest diameter (LD) for all CT image target lesions will be computed and reported as the baseline for the follow-up evaluations.
\end{itemize}
The RECIST 1.1 guideline consequently specifies how to classify the patient's response to a specific cancer therapy:

\begin{itemize}
\item A patient shows a complete response (CR) to the medical treatment if all identified target lesions disappear at the end-treatment CT imaging.
\item A patient shows a partial response (PR) if the target lesions (LD sum) are reduced by at least 30 \%. 
\item A patient shows a progressive disease (PD) if the LD sum increases by, at least, 20\% ;
\item A patient instead reports stable disease (SD) if no significant increase or decrease is observed on the target lesions. 
\end{itemize}

This work as well as the clinical trial to which it refers, provide results based on RECIST 1.1 guideline\cite{eisenhauer2009new}.

\subsection{The 2D-CNN Features Generative Model}

This block involves further processing of the collected ROIs to extract informative visual features. The proposed generative model utilizes an extended and appropriately configured version of 2D Cellular Nonlinear Networks (CNNs). The theory of Cellular Neural (or Nonlinear) Networks (CNNs) is briefly summarized.

The initial architecture of CNNs was proposed by L.O. Chua and L. Yang \cite{chua1988cellular}. CNNs can be described as a high-speed array of analog processors, referred to as "cells" \cite{chua1988cellular}. The fundamental unit of a CNN is the cell, and the processing of CNNs is governed by instructions provided by cloning templates \cite{chua1988cellular}. Each cell within a CNN can be viewed as a dynamical system arranged in a 2D or 3D topological structure. Interactions between CNN cells occur within a neighborhood defined by a heuristically determined radius \cite{chua1988cellular}.
Each CNN cell possesses an input, a state, and an output, with the state typically being mapped through a PieceWise Linear function. Stability analysis and considerations regarding the dynamics of CNNs have been documented in \cite{chua1988cellular, arena1996dynamics}.

It is worth noting that the utilized 2D-CNN belongs to the category of transient-response CNNs. These architectures perform the transformation of input data within a limited time frame, specifically during the transient stage. During this phase, each individual cell of the CNN dynamically evolves from its initial state along a trajectory that converges to the CNN's steady-state \cite{chua1988cellular, arena1996dynamics}. This allows us to extract intermediate 2D-CNN transformation features from the input data, which will be utilized as augmented generated features. In the CNN paradigm, the relationship between state, input, output, and neighborhood can be hypothesized in various ways, considering that CNNs are considered "Universal Machines" \cite{chua1992cnn, fortuna2001cellular}. Thus, we proceed with an extended mathematical representation of a generative model based on CNNs. The CNN can be considered as a system of cells (or non-linear neurons) defined on a normed-space $S_N$ (grid structure), which is a discrete subset of $R^n$ (generally $n\leq 3$) with distance function d: $S_N \rightarrow N$ (N is the set of positive integer numbers). Cells are indexed in a space-set $L_i$. Neighborhood function $N_r(.)$ of a k-th cell can be defined as follows:
\begin{equation}
\begin{split}
    &N_r:L_i\rightarrow L_i^\beta \\
    &N_r(k)=\{l|d(i,j) \le r\} 
\end{split}
\end{equation}
where $\beta$ depends on \textit{r} (neighborhood radius) and on space geometry of the grid. The CNNs can be implemented as a single layer or multi-layers so that the cell grid can be e.g., a planar array (with rectangular, square, octagonal geometry) or a k-dimensional array (usually $k\geq3$), generally considered and realized as a stack of k-dimensional arrays (layers). Therefore, CNN can be represented as a time-continuous - space-discrete system whose dynamic is well defined by the spatio-temporal evolution of the cells. The following mathematical model defines the  dynamic of an extended CNN cell:

%\begin{align}
\begin{equation}
\begin{split}
    \partial x_j/\partial t &= 
    g[x_j\ (t)] +\\
    &+\sum_{\substack{\gamma \in N_r(i)} } \aleph_{\vartheta_j}\left(x_j|_{\left(t-\tau,t\right]},y_\gamma{\ |}_{\left(t-\tau,t\right]};p_{\ j}^A\right) +\\
    &+\sum_{\gamma\in N_r\left(j\right)}{\mathcal{B}_{\varphi_j}\left(x_j|_{\left(t-\tau,t\right]},u_\gamma{\ |}_{\left(t-\tau,t\right]};p_{\ j}^B\right)} +\\ 
    &+\sum_{\gamma\in N_r\left(j\right)}{\mathbb{C}_{\rho_j}\left(x_j|_{\left(t-\tau,t\right]},x_\gamma{\ |}_{\left(t-\tau,t\right]};p_{\ j}^C\right)} +\\&+I_j\left(t\right) \\
\end{split}
\end{equation}
\begin{equation}
\begin{split}
y_j(t) = \psi(x_j|_{t-\tau,t}) 
\end{split}
\end{equation}

\begin{figure}

  \centering
  \centerline{\includegraphics[ width=\linewidth]{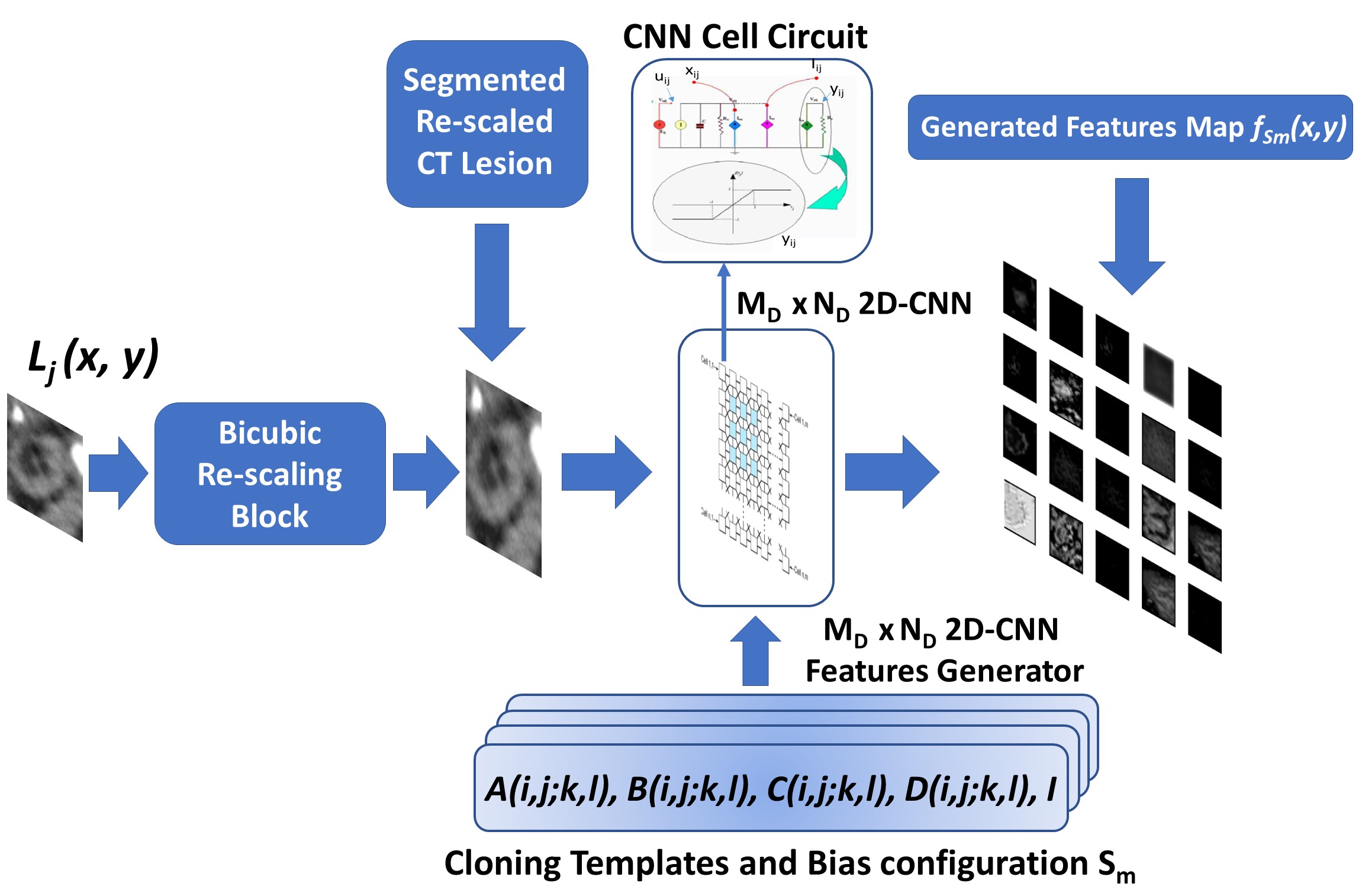}}
  \caption{\small The proposed 2D-CNN generative model }
  \label{fig:Img2} 
  \end{figure}

In Equation 2, the variables $x$, $y$, $u$, and $I_j$ represent the cell state, output, input, and bias, respectively. The indices $j$ and $\gamma$ refer to specific cells, while $g$ represents the local instantaneous feedback function. The function $N_r$ defines the neighborhood function, and $p^A$, $p^B$, and $p^C$ are arrays of configurable parameters specific to each cell. The notation $z|T$ denotes the restriction of function $z(\cdot)$ to the interval $T$ of its argument. The symbol $\aleph_{\vartheta_{j}}$ represents the neighborhood feedback functional, which is one of several applicable functions identified by the index $\vartheta_{j}$. Similarly, $\mathcal{B}{\varphi_j}$ denotes the input functional, and $\mathbb{C}{\rho_j}$ represents the cell-state functional. The term $I_j$ corresponds to an ad-hoc defined bias. The function $\psi$ describes the mathematical correlation between the cell state and its output.
Applying a linear re-mapping and extension of the model reported in Eqs. (2)-(3), the following CNN cell generative dynamical model  is proposed:

\begin{equation}
\begin{split}
%\MoveEqRight
C\frac{dx_{ij}\left(t\right)}{dt} &=-\frac{1}{R_x}x_{ij} +\\
&+\sum_{C\left(k,l\right)\in N_r\left(i,j\right)}{A\left(i,j;k,l\right)}y_{kl}\left(t\right) +\\
&+\sum_{C\left(k,l\right)\in N_r\left(i,j\right)}{B\left(i,j;k,l\right)}u_{kl}\left(t\right) +\\
&+\sum_{C\left(k,l\right)\in N_r\left(i,j\right)}{C\left(i,j;k,l\right)}x_{kl}\left(t\right) +\\
&+\sum_{C\left(k,l\right)\in N_r\left(i,j\right)}{D\left(i,j;k,l\right)}(y_{ij}\left(t\right),y_{kl}\left(t\right)) +\\
&+I \\ 
& 1\le i\le M,1\le j\le N 
\end{split}
\end{equation}

\begin{equation}
\begin{split}
    y_{ij}(t) &=\frac{1}{2}(|x_{ij}(t)+1|-|x_{ij}(t)-1|) \\
\end{split}
\end{equation}
\begin{equation}
\begin{split}
    N_r(i,j)&=\{C_r(k,l); (max{(|k-i|,|i-j|)}\le r)\} \\
    & (1\le k\le M,1\le l\le N))
\end{split}
\end{equation}
In Equations ($4$)-($6$), the term $Nr(i,j)$ represents the neighborhood of each CNN cell $C(i,j)$ with a radius of $r$. The variables $x_{ij}$, $y_{ij}$, $u_{ij}$, and $I$ denote the state, output, input, and bias of cell $C(i,j)$, respectively. The cloning templates $A(i,j;k,l)$, $B(i,j;k,l)$, $C(i,j;k,l)$, and $D(i,j;k,l)$ define the CNN processing, with $D(i,j;k,l)$ representing the extended non-linear space-invariant cloning template used in the proposed 2D-CNN model \cite{207503}. The configuration of the cloning templates determines the specific features generated by the CNN, as discussed in \cite{chua1992cnn, fortuna2001cellular, lee1996color}. The spatial dimensions of the CNN are defined by the setup (M, N) in Equations ($4$)-($6$), with $R_x$ representing an ad-hoc defined coefficient related to the hardware implementation of the CNN cell, set to 1.
In the implemented CNN generative features model, we input the segmented ROI lesion described in the previous section to both the input and state of the 2D-CNN. To match the input size of the downstream 2D-DNN ($M_D \times N_D$), we perform a preliminary bi-cubic rescaling of the segmented ROI to $M_D \times N_D$ size. Subsequently, an $M_D \times N_D$ 2D grid CNN is implemented. Each of the resized $M_D \times N_D$ segmented ROIs is fed into the input $u_{ij}$ and state $x_{ij}$ of the $M_D \times N_D$ 2D grid CNN, with each input and state of the CNN cells corresponding to the pixel values of the resized input visual lesion. In this case, the gray level intensity is used as the input for the CT ROI lesion, as it is a single-layer gray-level image. The workflow of the implemented CNN generative features model is illustrated in Figure \ref{fig:Img2}.
Let $L_{j} (x, y)$ denote the j-th visual segmented resized gray-level ROI image, which is fed as input ($u_{ij}$) and state ($x_{ij}$) to the $M_D \times N_D$ CNN grid. The image processing task, guided by the configured generative model according to Equations ($4$)-($6$), is performed. For each setup $S_{m}$ of the cloning templates $A(i,j;k,l)$, $B(i,j;k,l)$, $C(i,j;k,l)$, $D(i,j;k,l)$, and bias $I$ (generative model), a specific visual feature $f_{Sm}(x,y)$ of the input ROI $L_{j} (x, y)$ is generated. After conducting several tests, we have chosen to use $m=97$ different generative models, meaning 97 different configurations of $3 \times 3$ cloning template matrices and biases of size $1 \times 1$. This allows us to generate 97 augmented visual features for each segmented resized ROI $L_{j} (x, y)$.

\begin{figure}

  \centering
  \centerline{\includegraphics[width=\linewidth]{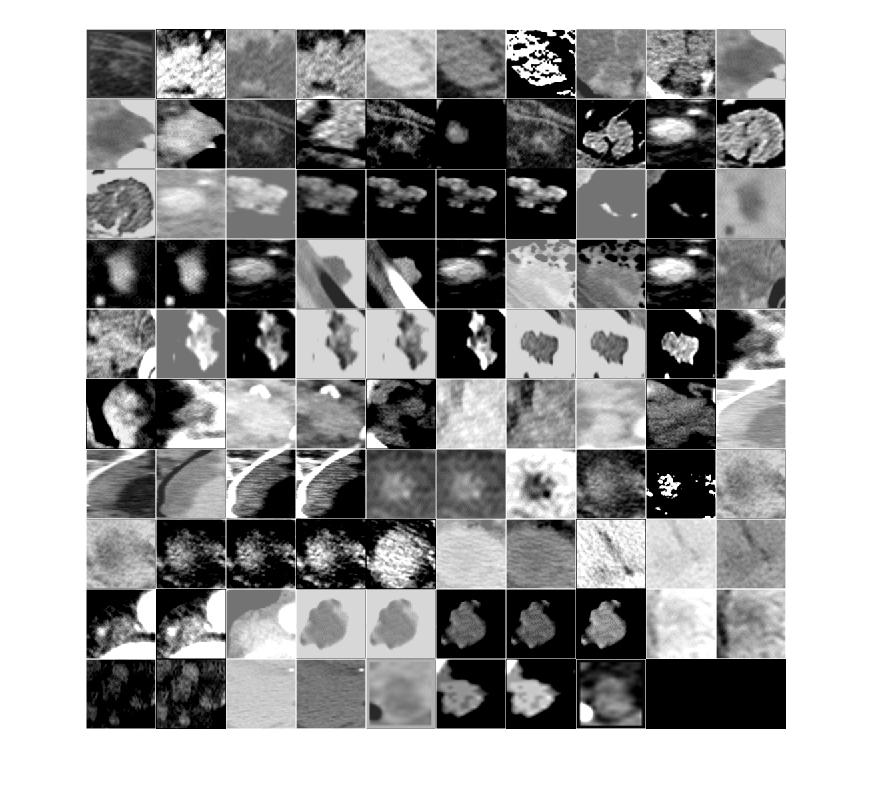}}
  \caption{\small A collection of the 2D-CNN generated features}
  \label{fig:Img3} 
  \end{figure}

Through an unsupervised approach that minimizes the loss of the downstream 2D-DNN classifier (maximizing the discrimination accuracy), we have selected the generative models. During the training process, a random-driven searching algorithm was employed to select the $3 \times 3$ cloning template and bias configurations $S_m$ that produced an improvement in the discriminatory performance of the 2D-DNN classifier. The designed generative models (cloning templates and biases) as well as the adopted transient configuration of the used 2D-CNN can be downloaded from ad-hoc designed web page\footnote{\url{https://iplab.dmi.unict.it/immunotherapy/}\label{note2}}. In Fig. \ref{fig:Img3} a collection of such 2D-CNN generated features is reported. The reported features are associated to different CT cancer native lesion types i.e. visceral, lymph-nodes, etc.. 

\subsection{The 2D-DNN Classifier with Decision System}

\begin{figure*}
  \centering
  \centerline{\includegraphics[width=0.9\linewidth]{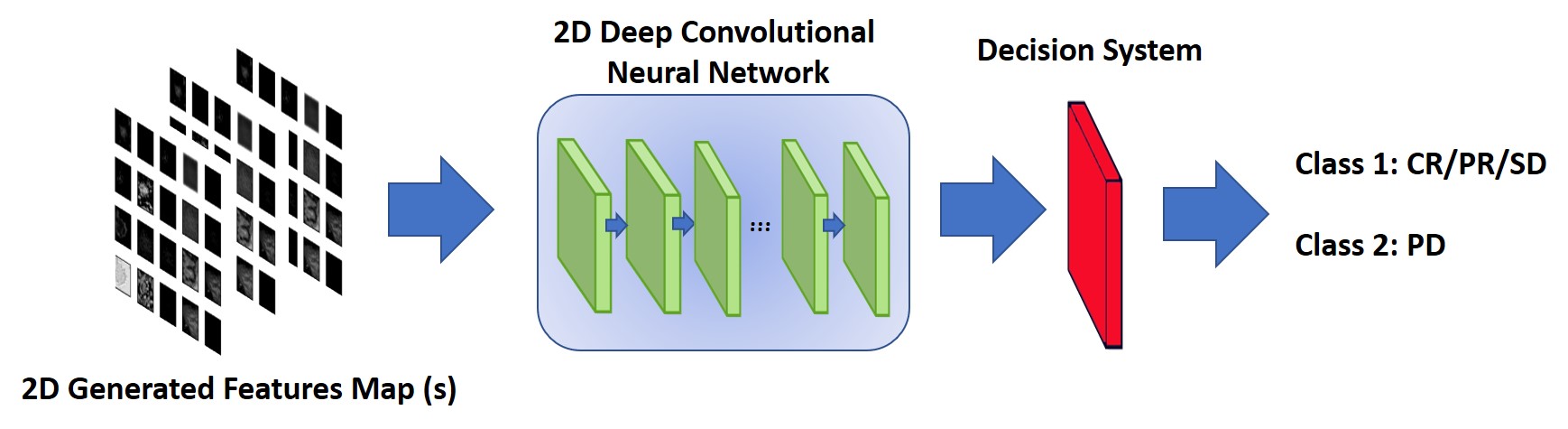}}
  \caption{\small The proposed immunotherapy outcome deep predictor. }
  \label{fig:Img4} 
  \end{figure*}

This block focuses on learning the 2D-CNN generated augmented features to accurately classify patients eligible for immunotherapy treatment. Fig. \ref{fig:Img4} illustrates the underlying architecture of the proposed deep model. Different 2D deep classifier backbones were investigated to assess their computational complexity on the implemented embedded hardware Point of Care and determine the ones that deliver the best predictive performance.
The following 2D deep classifier backbones were validated: ResNet-18, VGG-19, Xception, MobileNetV2, GoogleNet, and AlexNet \cite{LIU201711}. Additionally, a NasNetMobile-based 2D Deep Classifier was tested \cite{zoph2017learning}. For each patient, the target CT lesions (ROIs) that comply with the RECIST 1.1 guidelines are selected using the introduced semi-automatic segmentation block. The resized ROIs are then augmented using the proposed 2D-CNN generative engine. The generated augmented feature maps are subsequently classified by the 2D-DNN, which provides a probability estimation of belonging to class 1 (CR / PR: complete or partial response to immunotherapy treatment or SD: stable disease) or class 2 (PD: progressive disease).
In the last block, there is a Decision System depicted in Fig. \ref{fig:Img4}. This layer aims to collect the classifications of the individual augmented visual 2D-CNN generated features. As mentioned earlier, for each patient, a set of $m=97$ visual features is generated by the 2D-CNN generative model. Each of these feature images is then classified by the 2D-DNN as belonging to either class 1 or class 2. It is necessary to determine the main (statistically more representative) classification among the generated features associated with the patient whose CT visual lesion they refer to. The task of the Decision System, for each patient, is to determine the main 2D-DNN classification rate of the CNNs generated features. This predominant classification will be the definitive classification of the patient. Analytically, the Decision Layer will produce the following output:
\begin{equation}
    D_k = \xi_c(C_{DNN}(f_{Sm})); \ \ C_{DNN}(f_{Sm})=\{1,2\}
\end{equation}

Eq. (7) shows the mathematical model of the Decision Layer. The function $\xi_c$ will analyze the 2D-DNN classification $C_{DNN}$ rate of the $f_{Sm}$ generated features ($m=97$) related to the k-th patient, determining the class (1 or 2) that statistically is most represented. The output class of the Decision System ($\xi_c$)  becomes the definitive classification (immunotherapy outcome prediction) related to the analyzed patient. 

\section{Experimental Results}

We conducted a retrospective analysis using a dataset of 106 mUC cancer lesions extracted from chest-abdomen CT scans of patients with histologically confirmed bladder cancer who participated in clinical trials. These patients had bladder cancer that progressed after platinum-based chemotherapy and were subsequently treated with a PD-L1 ICIs immunotherapy agent in the second-line setting. All patients provided written informed consent, and the clinical trials were approved by the IRB "Catania 1 Ethical Committee" (Nr. D4191C00068 and MO29983).
For each patient, a chest-abdomen CT-scan was performed during the diagnostic phase for disease staging. The CT imaging was conducted using a GE multi-slice (64 slices) CT scanner, with a slice thickness of 2.5 mm. The clinical and personal history data of each patient complemented the CT scans. The aim of our proposed pipeline is to investigate the use of augmented CT image features at the time of bladder cancer diagnosis to predict the outcome of immunotherapy treatment.
Statistical information about the recruited dataset is as follows: approximately 30\% of the patients were under the age of 60, and 91\% of the patients were male, with the remaining 9\% being female. Among the subjects, about 33\% had lymph node metastases, while the remaining 67\% had various visceral metastatic lesions. Out of the 106 cases (target lesions), 43 showed a complete/partial response or disease stabilization following immunotherapy treatment (CR/PR/SD: Class 1), while 63 cases experienced disease progression despite anti-PD-L1 drug treatment (PD: Class 2).
The dataset was split as follows: 76 target lesions (28 of Class 1 and 48 of Class 2) were used for training and validation, and the remaining 30 CT images (15 of Class 1 and 15 of Class 2) were used as the test set. These lesions were augmented using the proposed generative model based on the 2D-CNN. Considering that for each CT image lesion, a generative engine of m=97 CNN models setup (3x3 cloning templates and 1x1 biases) was defined, the augmented dataset consists of a total of 7372 (76x97) image features (2716 images of Class 1 and 4656 images of Class 2) for the training and validation set. The test set comprises 2910 image features (1455 images for each class). 
The proposed pipeline was developed using MATLAB rev. 2019b with full toolboxes. The implementation was carried out on a server with an Intel 16-Core processor and an NVIDIA GeForce RTX 2080 GPU. We utilized the Deep Learning and custom matCNN MATLAB toolboxes for implementing the 2D-DNN and 2D-CNN generative models, respectively.

For all the tested 2D-DNN model backbones, we used the following learning parameters: a mini-batch size of 10, an initial learning rate of 3e-4, a maximum of 900 epochs, and the stochastic gradient descent with momentum (SGDM) algorithm as the learning optimizer.

We configured the 2D-CNN as an $M_d \times N_d$ grid according to the input image size of the tested 2D-DNN classifier, as specified in Table I. The 2D-CNN was set up with 3x3 cloning templates and 1x1 biases, which were made available for download on the web-page mentioned earlier\footref{note2}.
Table I presents the performance evaluation of the tested 2D-DNN architectures with 2D-CNN augmentation. The performance of the proposed pipeline was validated using classical metrics such as Accuracy, Sensitivity, and Specificity.

To evaluate the pipeline's performance, we considered "True Positive" as the correct classification of patients who showed a positive response to immunotherapy treatment (CR, PR, or SD) and were classified as belonging to Class 1. "True Negative" referred to patients who were classified as belonging to Class 2 and did not show a response to the immunotherapy drug (PD). "False Negative" and "False Positive" values were also computed accordingly.

Table I shows the performance results (on the test set) in terms of accuracy, sensitivity, and specificity using each selected 2D-DNN classifier with the augmented CT image dataset through the proposed 2D-CNN generative model. While some architectures achieved 100\% sensitivity, deep architectures demonstrated acceptable performance even in terms of specificity, which is preferred.

ResNet-18 and VGG-19 exhibited an interesting trade-off in classification performance. ResNet-18 achieved an accuracy of 93.33\%, sensitivity of 93.33\%, and specificity of 93.33\%. VGG-19 achieved an accuracy of 93.33\%, sensitivity of 100.00\%, and specificity of 86.66\%. All performance results are reported in Table I. The accuracy, sensitivity, and specificity values refer to the overall classification of the generated features performed by the Decision System as described in Eq. (7) ($\xi_c$ output).

The analysis of Table I confirms that despite the limited architectural complexity of the tested classifier backbones (due to hardware limitations of the embedded Point of Care), the overall prediction performances are highly promising.

 \begin{table}[ht!]
\caption{2D-DNN Performance Benchmark - 2D-CNN Dataset Augmentation Model}
    \begin{center}
    \resizebox{\columnwidth}{!}{
        \begin{tabular}{|c|c|c|c|c|}
        \hline
        \textbf{2D-Deep Classifier Backbone}&\multicolumn{4}{|c|}{\textbf{Metrics}} \\
        \cline{2-5} 
        \textbf{}& \textbf{\textit{Accuracy}}& \textbf{\textit{Sensitivity}}& \textbf{\textit{Specificity}}& 
        \textbf{\textit{Input Size}} \\
        \hline
        ResNet-18& $93.33 \%$& $93.33 \%$& $93.33 \%$& $224x224$ \\ \hline
        VGG-19&	$93.33 \%$& $100.00 \%$& $86.66 \%$& $224x224$ \\ \hline
        XCeption& $86.66 \%$& $93.33 \%$& $80.00 \%$& $299x299$ \\ \hline
        MobileNetV2&	$83.33 \%$& $100.00 \%$& $66.66 \%$& $224x224$\\ \hline
        GoogleNet&	$86.66 \%$& $86.66 \%$& $86.66 \%$& $224x224$ \\ \hline
        AlexNet&	$83.33 \%$& $80.00 \%$& $86.66 \%$& $227x227$ \\ \hline
        NasNetMobile&	$76.66 \%$& $80.00 \%$& $73.33 \%$& $224x224$ \\ \hline
        Previous\cite{rundo2019advanced}&	$86.05 \%$& $80.00 \%$& $89.29 \%$& $40x40$ \\
        \hline
        \end{tabular}
    %\label{table:tab1}
    }
    \end{center}
\end{table}

\begin{table}[ht!]
\caption{2D-DNN Performance Benchmark - Classical Dataset Augmentation Method}
\begin{center}
    \resizebox{\columnwidth}{!}{
        \begin{tabular}{|c|c|c|c|}
        \hline
        \textbf{Deep Classifier Backbone}&\multicolumn{3}{|c|}{\textbf{Metrics}} \\
        \cline{2-4} 
        \textbf{} & \textbf{\textit{Accuracy}}& \textbf{\textit{Sensitivity}}& \textbf{\textit{Specificity}} \\
        \hline
        ResNet-18& $86,66 \%$& $100.00 \%$& $78.66 \%$ \\ \hline
        VGG-19&	$73.33 \%$& $86.66 \%$& $60.00 \%$ \\ \hline
        3D-DenseNet&	$83.33 \%$& $86.66 \%$& $80.00 \%$ \\ \hline
        ResNet-101&	$76.66 \%$& $80.00 \%$& $73.33 \%$ \\
        \hline
        \end{tabular}
    }
    
\end{center}
\end{table}

Table II provides a comparison of the performance of the most effective architectures from Table I (ResNet-18 and VGG-19) without the 2D-CNN generative model, using traditional input augmentation approaches. The results demonstrate that the performance, in terms of overall performance, was significantly lower compared to the same architectures trained with a dataset augmented through 2D-CNN.

ResNet-18, without the 2D-CNN generative model, showed an improvement in sensitivity (100\% compared to 93.33\% with 2D-CNN) but a considerable decrease in specificity (78.66\% compared to 93.33\% with 2D-CNN), limiting the overall performance of the pipeline.

Similarly, VGG-19 without the 2D-CNN generative model exhibited significantly lower performance compared to when it was used with the upstream 2D-CNN generative model.

Even when compared to more complex architectures such as 3D-DenseNet (with a classification stage and 16 CT slices as input temporal depth) and 2D ResNet-101, which were tested without the use of the 2D-CNN generative model (as shown in Table II), the valid contribution of the 2D-CNN generative model in enhancing performance was evident.

These results highlight the effectiveness of the 2D-CNN generative model in improving the overall performance of the pipeline, surpassing the traditional input augmentation approaches and even outperforming more complex architectures.
\section{Conclusion and Discussion}

This study introduces an innovative pipeline for predicting immunotherapy treatment outcome in patients diagnosed with bladder cancer using augmented CT image features. Traditional approaches for predicting treatment response have limitations and invasiveness, making the development of non-invasive image-based biomarkers crucial. The pipeline focuses on learning image features from chest-abdomen CT scans of metastatic bladder cancer patients, leveraging the unique response patterns observed in ICIs treatment.

To overcome the challenge of limited labeled clinical data, an extended 2D-CNN generative model was employed and validated. The experimental results presented in Tables I and II demonstrate the promising performance of the proposed pipeline in terms of accuracy, sensitivity, and specificity. This improvement is attributed to the combination of a deep high-capacity classifier and the 2D-CNN augmented training set.

The pipeline successfully analyzes all RECIST 1.1 compliant lesions, indicating that it is not strongly influenced by lesion selection and is robust across different lesion types. The implementation of the pipeline is designed for the embedded STA1295 platform with OpenCV and YOCTO Linux OS.

Future studies aim to enhance the pipeline by incorporating a fully automatic segmentation block. An enhanced GradCAM-driven algorithm is being investigated, which integrates the RECIST 1.1 guideline to enable automatic CT ROI segmentation based on explainable visual features. Additionally, an adaptive Deep GAN architecture is being explored as a generative model.

Overall, the proposed pipeline shows promising potential for predicting immunotherapy treatment outcome in bladder cancer patients, and ongoing research aims to further enhance its capabilities.
%\section*{References}

%\begin{thebibliography}{00}
\bibliographystyle{IEEEtran.bst}
\bibliography{refs}
%\end{thebibliography}

\end{document}